\documentclass[twocolumn,aps,
preprintnumbers,amsmath,amssymb]{revtex4}
\usepackage{graphicx}
\newcommand{\beq}{\begin{equation}}
\newcommand{\eeq}{\end{equation}}

\begin{document}


\title{Crossover dark soliton dynamics in ultracold one-dimensional Bose gases}
\author{D.J. Frantzeskakis$^{1}$, P.G. Kevrekidis$^{2}$ and N.P. Proukakis$^{3}$}
\address{
$^{1}$ Department of Physics, University of Athens, Panepistimiopolis, Zografos, Athens 157 84, Greece 
\\
$^{2}$ Department of Mathematics and Statistics, University of Massachusetts, Amherst, Massachusetts 01003-4515, U.S.A.
\\
$^{3}$ School of Mathematics and Statistics, University of Newcastle, Merz Court, Newcastle NE1 7RU, United Kingdom
}

\begin{abstract}

Ultracold confined one-dimensional atomic gases are predicted to support dark soliton solutions 
arising from a nonlinear Schr\"{o}dinger equation of suitable nonlinearity.
In weakly-interacting (high density) gases, the nonlinearity is cubic, whereas an approximate 
model for describing the behaviour of strongly-interacting (low density) gases is one 
characterized by a quintic nonlinearity. We use an approximate analytical expression for 
the form of the nonlinearity in the intermediate regimes
to show that, near the crossover between the two different regimes, 
the soliton is predicted and numerically confirmed 
to oscillate at a frequency of $\sqrt{2/3}\Omega$, where $\Omega$ is the
harmonic trap frequency. 

\end{abstract}

\maketitle

\section{Introduction}

Dark solitons (DSs), the most fundamental nonlinear excitations of the 
one-dimensional defocusing nonlinear Schr\"{o}dinger (NLS) 
equation, have been studied in a broad range of physical systems. Apart from the theoretical work, 
experimental studies on DSs include their observation either as temporal pulses in optical fibers \cite{fiber}, 
or as spatial structures in bulk media and waveguides \cite{spatial} (see also \cite{book} for a review), the excitation 
of a nonpropagating kink in a parametrically-driven shallow liquid \cite{den1}, 
standing DSs in a discrete mechanical system \cite{den2}, high-frequency DSs in 
thin magnetic films \cite{magn}, and so on. 

Recently, DSs have attracted much attention in the physics of atomic Bose-Einstein condensates (BECs) 
\cite{review}, where dark matter-wave solitons have also been observed experimentally \cite{dark}. 
Dark solitons in BECs are known to be more robust in one-dimensional (1D) geometries and at very low
temperatures (a regime which is 
currently experimentally accessible \cite{q1db}). 
For that reason, the majority of theoretical studies on DSs in BECs have
for simplicity been performed in the framework of the one-dimensional (1D) {\it cubic} NLS equation, 
which in this context 
is referred to as the Gross-Pitaevskii equation; 
the latter, is the commonly adopted mean-field 
theoretic model describing ultracold {\it weakly-interacting} Bose gases in the absence of thermal or quantum fluctuations.
Many of the above mentioned theoretical studies have been devoted to the analysis of the 
dynamical properties of moving DS, such as their oscillations \cite{motion1,huang,sound,bkp,motion3} 
and sound emission \cite{huang,sound,motion3} in the presence of the external trapping potential. 
On the other hand, and in the same context of the 1D Bose systems, 
DSs have also been studied in the framework of a {\it quintic} NLS equation \cite{kolom1,fpk,kavoulakis,bkp}, 
a long-wavelength model which has been proposed \cite{kolom1} for the opposite limit of {\it strong interatomic coupling} 
\cite{dunjko,petrov}; in this case, the collisional properties of the bosonic atoms are significantly modified, 
with the interacting bosonic gas behaving like a system of free fermions \cite{tg}. 
Such, so-called, Tonks-Girardeau (TG) gases have recently been observed experimentally as well \cite{bl}. 


Regarding the dynamical features of the moving DSs in trapped 1D Bose gases, 
the above works revealed that, in the absence of other dissipative mechanisms, 
the soliton oscillates in the trap with a frequency which differs between the weakly and strongly interacting regimes. 
In particular, in the presence of a harmonic confining potential of frequency $\Omega$, 
the study of the cubic (quintic) NLS predicts such an oscillation frequency to be $\alpha \Omega$, 
where $\alpha =  1/\sqrt{2}$ ($\alpha=1$) for the weakly \cite{motion1,huang,sound,bkp,motion3} 
(strongly \cite{fpk,bkp}) interacting Bose gas. 
The latter result for the strongly interacting case 
is identical to the corresponding one obtained by a full many-body calculation \cite{busch}.

The transition between the weakly and strongly interacting regimes is usually characterized by
a single parameter, denoted by $\gamma$, quantifying the 
ratio of the average interaction energy to the kinetic energy calculated with mean field theory \cite{petrov}.
This parameter varies smoothly as the interatomic coupling is increased from values $\gamma \ll 1$ 
(weakly interacting 
regime), to $\gamma \gg 1$ (strongly interacting regime); thus,
an approximate ``crossover regime'' can be identified around $\gamma \sim O(1)$, as also attained experimentally
\cite{ool}. 
The value of $\gamma$ can be controlled experimentally by various independent parameters,
such as scattering length, transverse confinement, density, or even modification of the effective mass of the system.


Motivated by the investigation of the DS dynamics in 
the two limiting regimes, in the present work we obtain analytical results, which are 
confirmed by numerical simulations, for
the oscillation frequency of a DS in the {\it crossover} regime for a purely 1D system. 
In particular, we show that for $\gamma \sim O(1)$ the DS oscillates with a frequency $\sqrt{2/3} \Omega$, 
which is higher than the one ($\sqrt{1/2} \Omega$) pertaining to the weakly interacting Bose gas. 
Such an increase in the soliton oscillation frequency could serve as an additional diagnostic test for the 
deviation from pure bosonic mean field behaviour. 

The paper is organized as follows: In section II we present the generalized 
NLS model proper and find its ground state, as well as its linear (sound waves) and nonlinear (dark solitons) excitations. 
Section III is devoted to the analytical derivation of the DS oscillation frequency and the discussion of relevant 
numerical simulations. Finally, in section IV we summarize our findings.

\section{The model and its analytical consideration}

\subsection{The generalized NLS equation}

Importantly for our present analysis, 
the parameter $\gamma$ separating the regimes of weak and strong interatomic coupling
can be re-expressed, for a given system configuration,
 in terms of the inverse ratio of the system density to some critical density. 
This enables us to perform a unified analysis of all regimes by means of a NLS equation 
with a generalized nonlinearity, for the parameter 
$\psi$ connected to the density $n = |\psi|^2$ of an ultracold confined 1D Bose gas.
This equation takes the form 
\beq
i\hbar \frac{\partial \psi}{\partial t}=
\left[-\frac{\hbar^{2}}{2m}\frac{\partial^{2}}{\partial x^{2}}
+ V_{\rm ext}(x) +
\Phi (n) \right]\psi,
\label{gpetg}
\eeq
where 
$m$ is the atomic mass and $V_{\rm ext}(x)= (1/2)m \omega_{x}^2 x^2$ is the external trapping potential 
($\omega_{x}$ being the axial confining frequency).  

The exact form of the nonlinearity $\Phi(n)$ valid in both limiting regimes and the
crossover region is well-known in the homogeneous hydrodynamic limit. While the functional 
dependence of $\Phi(n)$ on $\gamma$ (and its analytical asymptotics) are known \cite{LL}, 
its precise values in the ``crossover region'' can only be evaluated numerically. 
Such intermediate values have been tabulated by Dunjko et al. in \cite{dunjko}, 
and subsequently discussed by various authors in the local density approximation, 
see e.g. \cite{santos,jb}. Since we are interested in this ``crossover region'', 
we should thus use an approximate expression for the nonlinearity, which captures 
both limiting regimes exactly and provides a good approximation for intermediate 
values. At the same time, however, we are constrained in our present work by the 
need for a relatively simple expression which will enable rather involved analytical 
work to be carried out. For our purpose, it is thus sufficient to use
%
a somewhat simplified generalized nonlinearity $\Phi(n)$ of the form
\beq
\Phi(n)= \frac{\pi^{2} \hbar^{2}}{2m} \frac{n^{2}}{1+n/\tilde{n}_{c}},  
\label{Phi}
\eeq
%
%
which nonetheless corresponds to 
a fairly good analytical approximation \cite{jb,luisjoachim} 
to the exact nonlinearity 
\cite{dunjko}. 
In this notation, the critical density approximately marking the crossover region
is given by $\tilde{n}_{c} = 4 / (\pi^{2} a_{1D})$,
where $a_{1D}$ is the effective 1D ``scattering length''. This is 
defined in terms of the usual three-dimensional 
s-wave scattering length $a_{3D}$ via $a_{1D}=l_{\perp}^{2}/a_{3D}$, where
$l_{\perp} = \sqrt{\hbar/ m \omega_{\perp}}$ is the harmonic oscillator length 
in the transverse direction.

Equation (\ref{gpetg}) is well-documented in the high-density limit $n \gg \tilde{n}_{c}$
(corresponding to $\gamma \ll 1$). In this case,
the parameter $\psi$ describes the ``wavefunction'' 
of a weakly-interacting 1D Bose gas under harmonic confinement, for which 
Eq. (\ref{gpetg}) reduces to the 1D Gross-Pitaevskii equation, 
\beq
i\hbar \frac{\partial \psi}{\partial t}=
\left[-\frac{\hbar^{2}}{2m}\frac{\partial^{2}}{\partial x^{2}}
+ V_{\rm ext}(x) +
\frac{4 \pi \hbar^{2}}{m} \left( \frac{a_{3D}}{2 \pi l_{\perp}^{2}} \right) |\psi|^{2} \right]\psi.
\label{usual}
\eeq
%

In the opposite limit of strong interatomic coupling, 
which, rather counter-intuitively, corresponds to the low density limit $n \ll \tilde{n}_{c}$,
the parameter $\psi$ 
satisfies the following quintic NLS equation,
\beq
i\hbar \frac{\partial \psi}{\partial t}=
\left[-\frac{\hbar^{2}}{2m}\frac{\partial^{2}}{\partial x^{2}}
+ V_{\rm ext}(x) +
\frac{\pi^{2} \hbar^{2}}{2m} |\psi|^{4} \right]\psi,
\label{kolom}
\eeq
which has been derived by various different methods in
\cite{kolom1,qnls1}. 
While the validity of Eq. (\ref{kolom}) 
to discuss coherence properties of strongly-interacting 1D Bose gases has been questioned 
\cite{girardeau_bad}, 
the corresponding 
hydrodynamic equations for the density $n$ and the phase $\phi$
(or the atomic velocity $v \equiv \partial_{x} \phi$) 
arising from this equation under the Madelung transformation $\psi = \sqrt{n} \exp(i \phi)$ are well-documented in the
context of the local density approximation \cite{dunjko,santos}. 
An equation of the form (\ref{kolom}), which explictly includes the so-called  
``quantum pressure term'' $(\hbar^{2}/2m)\partial^{2} \psi /\partial x^{2}$,
should however only be valid for density variations which occur on a lengthscale which is larger
than the Fermi healing length $\xi \equiv 1/(\pi n_{\rm p})$, where $n_{\rm p}=|\psi(0)|^{2}$ 
is the peak density of the gas at the trap center. 
To avoid such potential complications, the physical analysis presented in this paper 
is only concerned with the limit of shallow DSs, for which there is only a very slow 
density variation within the Fermi healing length. 

Measuring the variables $x$ and $t$, and the density $|\psi|^2$, 
in units of the Fermi healing length $\xi$, 
the time $\tau_{0}=\hbar/m \xi^{2}$, and the density $\sqrt{2} n_{\rm p}$ respectively, 
Eq. (\ref{gpetg}) can be rewritten in the following dimensionless form,
\begin{equation}
iu_{t}=-\frac{1}{2}u_{xx}+V(x)u+g(n)u,
\label{gpe}
\end{equation}
where the subscripts denote partial differentiation. In view of the above scalings, 
the normalized confining potential becomes $V(x)=(1/2)(\xi/l_x)^{4}x^{2}$,
where $l_{x} = \sqrt{\hbar/ m \omega_{x}}$ is the harmonic oscillator length 
in the axial direction. As the parameter $\xi/l_x$ is apparently small, it is convenient to define the small parameter
$\epsilon \equiv \Omega^{-2/3}(\xi/l)^{4/3}$ where $\Omega$ is a parameter of order $O(1)$, 
that will be used in the perturbation analysis to follow. This way, the external potential 
is actually a function of the slow variable $X \equiv \epsilon^{3/2}x$, and has the form $V(X)=(1/2)\Omega^{2}X^{2}$, 
where $\Omega$ expresses the trap frequency. Finally, the nonlinearity function $g(n)$ 
(with $n=|u|^{2}$ being the normalized density) becomes,
\beq
g(n)=\frac{n^2}{1+n/n_{c}}, \,\,\,\ \  {\rm where} \,\,\,\, n_{c}=\frac{2 \sqrt{2}}{\pi^2}\frac{1}{a_{1D}n_{\rm p}},
\label{fn}
\eeq

The weakly and strongly interacting limits discussed above respectively arise when the dimensionless parameter 
$\gamma = 2/ (n a_{1D})$ \cite{petrov} obeys $\gamma \ll 1$ or $\gamma \gg 1$. 
In our present notation, $n/n_c=(\pi^{2}/2)\gamma^{-1}$, so that our approximate 
crossover region actually corresponds to a value $\gamma = 2/\pi^{2} \approx 0.2$. 
For such a relatively small value of $\gamma$, it is still reasonable 
to use Eqs. (\ref{gpetg})-(\ref{Phi}) [or 
Eqs. (\ref{gpe})-(\ref{fn})] 
to describe this physical system.

Studies of DSs based on generalized 1D NLS equations first appeared 
in the literature for homogeneous systems 
to deal with 
saturable nonlinearities appearing in the context of nonlinear optics \cite{book}; 
in this case, the nonlinearity becomes cubic at low densities, rather than at high densities which 
occurs for ultracold pure 1D Bose gases.
More recently, DSs in generalized 1D NLS equations were also considered in the BEC context, but only as effective theories for
weakly-interacting elongated 3D condensates. 
These equations contained either a cubic-quintic nonlinearity with constant coefficients \cite{muryshev}, or a generalized non-polynomial nonlinearity depending explicitly on the trap aspect ratio \cite{salasnich,salasnich_new}. 
The aim of this work is rather distinct, namely to calculate the soliton oscillation frequency in the crossover between weakly- and strongly interacting 1D Bose gases.
The validity of our work is thus restricted to the pure 1D regime, with the precise experimental conditions needed for such a crossover discussed in detail in \cite{dunjko,menotti}.


Below we apply the reductive perturbation method (see, e.g., \cite{rpm,asano}), valid in the limit of shallow solitons, to 
analytically obtain the soliton oscillation frequency close to the critical density $n_{c}$, which approximately
marks the crossover between the regimes of weak and strong interatomic coupling. 


\subsection{Ground state, linear and nonlinear excitations}

Starting off from the generalized NLS Eq. (\ref{gpe}), 
we use the Madelung transformation, $u=\sqrt{n}\exp(i\phi)$, 
to obtain the following set of hydrodynamic equations, 
%
\begin{eqnarray}
&& n_{t}+(n\phi_{x})_{x}=0, \label{h1} \\
&& \phi_{t}+g(n)+\frac{1}{2}\phi_{x}^{2}-\frac{1}{2}n^{-1/2}(n^{1/2})_{xx}
+V(X)=0. \label{h2}
\end{eqnarray}
The above equations 
are similar to the ones that have been employed to discuss the crossover from TG to BEC regime 
\cite{dunjko,santos}.
The ground state of the system can be obtained upon assuming that the atomic velocity 
$v \equiv \phi_{x}=0$ (i.e., no flow in the system) and $\phi_{t}=-\mu_{0}$ 
(dimensionless chemical potential). Then, as Eq. (\ref{h1}) implies that $n=n_{0}$ is 
time-independent in the ground state, we assume that $n_{0}=n_{0}(X)$. 
Thus, to leading order in $\epsilon$ [i.e., to $O(1)$] Eq. (\ref{h2}) yields, 
\begin{equation}
g(n_{0})=\mu_{0}-V(X),
\label{TF}
\end{equation}
in the region where $\mu_{0}>V(X)$ and $n_{0}=0$ outside. Equation (\ref{TF}) determines the 
density profile in the so-called Thomas-Fermi (TF) approximation; note that for the typical 
case of the harmonic trap, e.g., $V(X)=(1/2)\Omega^{2}X^{2}$, Eq. (\ref{TF}) recovers the 
well-known result that the density profile is parabolic, with $n_{0}(X)=n_{c}^{-1}\left[\mu_{0}-V(X)\right]$, 
in the regime $n \gg n_{c}$ and elliptic, with $n_{0}(X)=\sqrt{\mu_{0}-V(X)}$, in the regime
$n \ll n_{c}$. 
Also, it is noticed that from Eq. (\ref{TF}), and for the harmonic trap under consideration, 
the axial size of the gas is $2L$, where $L=\sqrt{2\mu_{0}}/\Omega$ is the TF radius. 

We now consider the propagation of small-amplitude linear excitations (e.g., sound waves) 
of the ground state, by seeking solutions of Eqs. (\ref{h1})-(\ref{h2}) of the form
$n=n_{0}(X)+\epsilon\tilde{n}(x,t)$ and $\phi=-\mu_{0}t+\epsilon \tilde{\phi}(x,t)$, 
%
%
where the functions $\tilde{n}$ and $\tilde{\phi}$ describe the linear 
excitations. Inserting this ansatz 
into Eqs. (\ref{h1})-(\ref{h2}), to order 
$O(1)$ we recover the TF approximation, while to order $O(\epsilon)$ we obtain 
a system of linear equations for the linear excitations. 
%
%
Assuming plane wave solutions of this system, i.e., 
$(\tilde{n}, \tilde{\phi}) \sim \exp[i(kx-\omega t)]$, 
we readily obtain the 
dispersion relation $\omega^2=\dot{g}_{0}n_{0}k^{2}+ k^{4}/4$, 
%
%
where $\dot{g}_{0} \equiv (dg/dn)|_{n=n_{0}}$. 
This dispersion relation has the form of 
a Bogoliubov-type excitation spectrum, but with the excitation frequency 
$\omega$ being a function of the slow variable $X$. The speed of sound is local, due to 
the presence of the external potential,  
and is given by
\begin{equation}
C = \sqrt{ \dot{g}_{0}n_{0}},
\label{cs}
\end{equation}
%
Note that Eq. (\ref{cs}) shows that the speed of sound is given by $C=\sqrt{n_{c}n_{0}}$
for $n \gg n_{c}$, and 
$C=\sqrt{2}n_{0}$ in the opposite regime $n \ll n_{c}$.


 
Next we analyze the evolution of the {\it nonlinear} excitations 
on top of the ground state, employing the reductive perturbation method \cite{rpm,asano} 
(see also \cite{huang} and \cite{fpk} for relevant studies in Bose gases). 
As the system of Eqs. (\ref{h1})-(\ref{h2}) is inhomogeneous, we 
introduce a new slow time-variable $T=\epsilon^{1/2}\left( t-\int_{0}^{x} C^{-1}(x')dx'\right)$, 
and the following asymptotic expansions for the 
density $n$ and phase $\phi$,
\begin{eqnarray}
n&=&n_{0}(X)+ \epsilon n_{1}(X,T)+\epsilon^{2} n_{2}(X,T)+\cdots, \nonumber \\
\phi&=&-\mu_{0}t+ \epsilon^{1/2} \phi_{1}(X,T)+\epsilon^{3/2} 
\phi_{2}(X,T)+\cdots.
\label{ae}
\end{eqnarray}
Substituting the expansions (\ref{ae}) into Eqs. (\ref{h1})-(\ref{h2}), we 
obtain the following results: First, to order $O(1)$, Eq. (\ref{h2}) leads to 
the TF approximation [see Eq. (\ref{TF})]. Then, 
to the first-order of approximation in $\epsilon$ [i.e., 
to orders $O(\epsilon)$ and $O(\epsilon^{3/2})$], 
Eqs. (\ref{h2}) and (\ref{h1}) yield the equation, 
%
%
%
\begin{equation}
\phi_{1}(X,T)=-\dot{g}_{0}(X)\int_{0}^{T} n_{1}(X,T')dT',
\label{r}
\end{equation}
connecting the unknown functions $n_{1}$ and $\phi_{1}$.
%
Finally, to the second order of approximation [to order $O(\epsilon^{2})$ and $O(\epsilon^{5/2})$], 
Eqs. (\ref{h2}) and (\ref{h1}) 
%
%
lead to the following nonlinear evolution equation for $n_{1}$, 
%
\begin{eqnarray}
%
%
&&n_{1X}-\frac{\left(3\dot{g}_{0}+n_{0}\ddot{g}_{0} \right)}{2C^{3}} n_{1}n_{1T}+\frac{1}{8C^{5}}n_{1TTT} \nonumber \\ 
&&= -\frac{d}{dX}\left[ \ln \left(|C|\dot{g}_{0} \right)^{1/2} \right] n_{1}, 
\label{kdv1}
\end{eqnarray}
where $\ddot{g}_{0} \equiv (d^{2}g/dn^{2})|_{n=n_{0}}$. 
Equation (\ref{kdv1}) has the form of a Korteweg-deVries (KdV) equation with variable coefficients, which has been 
used in the past to describe shallow water-waves over variable depth, or ion-acoustic 
solitons in inhomogeneous plasmas \cite{asano}. Moreover, such KdV equations have recently been used to 
analyze the dynamics of DSs in Bose gases both in the weakly-interacting \cite{huang} 
and the strongly-interacting \cite{fpk} regimes. 

As the inhomogeneity-induced dynamics of the KdV solitons has been studied analytically 
in the past \cite{karpman}, we may employ these results to analyze the coherent evolution 
of DSs in the Bose gas under consideration. Thus, introducing the transformations 
$\chi=\int(8C)^{-5}dX$ and  
$n_{1}=(3/2)\left(3\dot{g}_{0}+n_{0}\ddot{g}_{0} \right)^{-1}C^{-2}\upsilon(\chi,T)$, 
we first put Eq. (\ref{kdv1}) into the form,
\begin{equation}
\upsilon_{\chi}-6\upsilon\upsilon_{T}+\upsilon_{TTT}=\lambda(\chi)\upsilon,
\label{kdv2}
\end{equation}
where $\lambda(\chi) \equiv (d/d\chi) \ln \left[ n_{0}^{3/4}\dot{g}_{0}^{1/4} \left(3\dot{g}_{0}+n_{0}\ddot{g}_{0} \right) \right]$.
%
%
In the case $\lambda=0$, i.e., for a homogeneous gas with $n_{0}(X)=n_{\rm p}={\rm const.}$, Eq. (\ref{kdv2}) is the 
completely integrable KdV equation, which possesses a single-soliton solution of the following form \cite{abl},
\begin{equation}
\upsilon=-2\kappa^{2}{\rm sech}^{2}Z, \,\,\ Z=\kappa\left[T-
\zeta(\chi)\right],
\label{sol}
\end{equation}
where $\zeta(\chi)=4\kappa^{2}\chi+\zeta_{0}$ is the soliton center (with
$d\zeta/d\chi=4\kappa^{2}$ being the soliton velocity in the $T$-$\chi$ 
reference frame), while $\kappa$ and $\zeta_{0}$ are arbitrary constants presenting 
the soliton's amplitude (as well as inverse temporal width) and initial position respectively. 
Equation (\ref{sol}) describes a density notch on the backround density $n_{\rm p}$, 
with a phase jump across it [see Eq. (\ref{r}), which implies that $\phi_{1}\sim \tanh Z$] 
and, thus, it represents an approximate DS solution of Eq. (\ref{gpe}).

On the other hand, in the general case of the inhomogeneous gas [i.e., in the presence of $V(X)$], 
soliton dynamics can still be studied analytically, provided that the
right-hand side of Eq. (\ref{kdv2}) can be treated as a small perturbation. 
As $\lambda(\chi)$ is apparently proportional to the density gradient, such a perturbative study 
is relevant in regions of small density gradients (e.g., near the trap center for 
a harmonic trapping potential), which is consistent with the use of the local density approximation.
In this case, employing the adiabatic perturbation theory for solitons \cite{ps}, 
we may seek for the soliton solution of Eq. (\ref{kdv2}) 
in the form of Eq. (\ref{sol}), but with the soliton parameters $\kappa$ and 
$\zeta$ being now unknown functions of $\chi$. 
The respective 
evolution equations for the soliton's amplitude 
and center  
%
%
can be solved analytically \cite{karpman} and the results, expressed in terms of the slow variable $X$, read:
\begin{eqnarray}
&&\kappa(X)=\kappa(0)\left(\frac{\sigma(X)}{\sigma(0)}\right)^{2/3}, 
\label{kx1} \\ 
&&\zeta(X)=\frac{1}{2}\kappa^{2}(0)\int_{0}^{X} C^{-5}(X') 
\left(\frac{\sigma(X')}{\sigma(0)}\right)^{4/3}dX' \nonumber \\
&&+\frac{1}{2\kappa(0)}\left[1-\left(\frac{\sigma(X)}{\sigma(0)}\right)^{-2/3}\right],
\label{kx2}
\end{eqnarray}
where $\kappa(0)$ and $\sigma(0)$ are the values of the respective functions 
at $X=0$. 

Note that the above procedure is general, and does not rely on the ratio of the parameter $n/n_{c}$,
although it has been shown to yield the correct results in both limits $n \gg n_{c}$ \cite{huang} and
$n \ll n_{c}$ \cite{fpk}. In this work, we use the above
general results for the evolution of the soliton parameters, 
to derive the equation of motion of the DS and 
find its oscillation frequency in the ``nonlinearity crossover regime'' $n \approx n_{c}$.


\section{Soliton Oscillation Frequency}
 
Confining ourselves to the ``crossover regime'' $n \approx n_{c}$, 
we may use a Taylor expansion of the function $g(n)$ around $n=n_{c}$, 
namely $g(n) \approx g(n_{c})+\dot{g}(n_{c})(n-n_{c})+\cdots$, 
to derive the approximate expression $g_{0} \approx n_{c}(3n_{0}-n_{c})/4$; 
the latter, along with the TF approximation [see Eq. (\ref{TF})], leads to the following density profile, 
\beq
n_{0}=\frac{4\sqrt{\mu_{0}}}{3\delta}\left[1+\frac{1}{4}\delta^{2}-\left(\frac{X}{L}\right)^{2}\right],
\label{n0c}
\eeq
where $\delta \equiv n_{c}/\sqrt{\mu_{0}}$ (it is reminded that $L \equiv \sqrt{2\mu_{0}}/\Omega$ 
defines the axial size of the gas). 
Based on Eq. (\ref{n0c}), 
it is now possible to derive the equation of 
motion of the DS as follows. First, we find the soliton phase, which, to order 
$O(\epsilon^{3/2})$, reads,
\beq
Z=\epsilon^{1/2} \kappa(X) \left[ t-\int \frac{dX'}{C(X')}-\epsilon 
\frac{9\sqrt{3/2}\kappa^{2}(0)}{\mu_{0}^{2}\Omega \left(4+\delta^{2}\right)^{5/2}} 
\left(\frac{X}{L}\right) \right].
\label{solph}
\eeq
Then, looking along the characteristic lines of soliton motion, it is possible to show that 
the position of the soliton satisfies the following equation of motion,
\begin{equation}
\frac{dX}{dt}=\frac{C}{1+\epsilon (9\sqrt{3}/2)\kappa^{2}(0)(4\mu+\delta^{2})^{-5/2}C}.
\label{emg}
\end{equation}
For sufficiently small $\epsilon$ the second term in the denominator can be neglected; 
in this case, Eq. (\ref{emg}) shows that the velocity of sufficiently shallow DSs 
is approximately the same as the speed of sound given by Eq. (\ref{cs}), 
i.e., $dX/dt \approx \sqrt{\dot{g}_{0}n_{0}}$. 
Thus, Eq. (\ref{emg}) can be approximated by the separable differential equation, 
\beq
\frac{n_{0}+n_{c}}{n_{0}\sqrt{2n_{c}^{2}+n_{0}n_{c}}}dX=dt,
\label{sep}
\eeq
which can readily be integrated. In particular, taking into account Eq. (\ref{n0c}), 
we find that Eq. (\ref{sep}) leads to the following result:
\beq
\frac{3}{4 \sqrt{13}} h(\delta) \ln w(X)
+\arcsin \left(\frac{X}{\tilde{L}}\right)=\left(\Omega\sqrt{\frac{2}{3}}\right)t,
\label{emi}
\eeq
where $h(\delta)=\delta\left(1+\frac{1}{4}\delta^{2}\right)^{-1/2}$, 
\beq
w(X)=\frac{\sqrt{\tilde{L}^{2}-X^{2}}+\frac{\sqrt{13}}{2} h(\delta)X}
{\sqrt{\tilde{L}^{2}-X^{2}}-\frac{\sqrt{13}}{2} h(\delta)X},
\label{H}
\eeq
and $\tilde{L} \equiv L \sqrt{1+\frac{7}{2}\delta^{2}}$. It can be seen that in the regions of 
small density gradient where $X/L$ is sufficiently small [which is consistent with the assumption 
that the perturbation $\sim \lambda(\chi)$ in the KdV Eq. (\ref{kdv2}) is weak], 
and for sufficiently small values of the parameter $\delta$, the first term on 
the left-hand side of Eq. (\ref{emi}) can safely be neglected \cite{comment1}. 
Then, it is readily seen that Eq. (\ref{emi}) 
is reduced to the following equation,
\begin{equation}
X= L \sin \left(\sqrt{\frac{2}{3}} \Omega t \right).
\label{m}
\end{equation}
Equation (\ref{m}) demonstrates that a shallow DS displays an oscillatory motion 
in the harmonic trap $V(X)=(1/2)\Omega^{2}X^{2}$ in the ``nonlinearity crossover regime''
$n \approx n_{c}$ of Eq. (2), with an oscillation frequency $\Omega_{\rm osc}$ given by
\beq
\Omega_{\rm osc}=\Omega \sqrt{\frac{2}{3}}. 
\label{of}
\eeq

\begin{figure}[tbp]
\includegraphics[width=8.5cm]{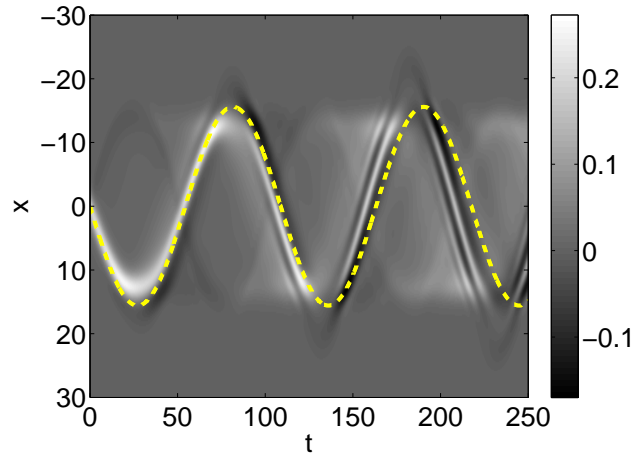}
\caption{Spatio-temporal evolution of the reduced condensate density
(the ground state density minus the actual density) for the NLS
equation with $g(n)=n^2/(1+n)$ and $V(x)=\Omega^2 x^2/2$, with $\Omega=0.0707$.
The dashed line shows the theoretical prediction which agrees well with 
the full numerical result over (at least) the first oscillation period.
}
\label{fig1}
\end{figure}

This finding has also been verified through appropriately crafted numerical
experiments. As a typical example, we show, in particular, 
the evolution of a shallow gray
soliton (originally located at the origin) 
with initial speed $C=0.9$ on the background of a potential
$V(x)=\Omega^2 x^2/2$, with $\Omega=0.0707$; $n_c$ was chosen
to be $1$ in this case. Figure \ref{fig1} shows the evolution of
the space-time contour plot of the reduced density 
(the ground state density minus the actual density) for the NLS
equation, alongside our theoretical prediction of Eq. (\ref{m}).
It is clear that during the first period, where the solitary wave
does not significantly interact with the emitted ``radiation'', 
the agreement between the theoretical
prediction and the numerical results is very good. Subsequently, 
as expected given the above interaction, this agreement deteriorates.
The approximate initial condition is produced by imposing a NLS
gray soliton on top of the ground state of the system for the nonlinearity
of interest. Finally, we note in passing that our analytical
considerations are, strictly speaking, valid away from the turning
points (where the wave speed vanishes). However, our numerical results,
as well as alternative approaches such as the ones of \cite{motion3} and \cite{bkp},
illustrate that the range of validity of our results is, in fact,
wider than what may be expected based on the mathematical 
limitations of the method.

Note that the established values of the oscillation frequencies in the two limiting regimes
of interatomic coupling 
can also be found in the framework of the presented analysis, upon utilizing Eq. 
(\ref{sep}) in the relevant limits $n/n_{c} \gg 1$ and $n/n_{c} \ll 1$ and using the 
respective density profiles.
In the limit $n \gg n_{c}$ coresponding, e.g. to a weakly-interacting Bose-Einstein condensate,
 the oscillation frequency is 
$\Omega_{\rm osc}=\Omega / \sqrt{2}$ (or $\Omega \sqrt{2/4}$), whereas, in the opposite regime
$n \ll n_{c}$ 
it has recently been found \cite{busch,fpk,bkp} 
that the oscillation frequency is 
$\Omega_{\rm osc}=\Omega$ (or $\Omega \sqrt{2/2}$).
The oscillation frequency in Eq. (\ref{of}) is thus predicted to lie between 
the above mentioned limiting cases. This suggestes a continuous change in the predicted
soliton oscillation frequency from the regime $n \gg n_{c}$ to $n \ll n_{c}$.

Assuming that the temperature is low enough for the soliton to perform many oscillations before decaying due to
additional dissipative mechanisms excluded from the NLS equation, an observation of change in the
oscillation frequency can be treated as a possible diagnostic tool 
of the system being in a particular interaction regime.
A relevant experiment might thus be to create a Bose gas of a certain unknown 
interaction strength, and phase imprint a DS in such a system. Measurement of the oscillation
frequency of this soliton will then provide important information on the system parameter regime,
with any deviation from the oscillation frequency of $\Omega/\sqrt{2}$ 
denoting a regime of 
sufficiently strong correlations which, in turn, indicate
that the weak-interaction
model is no longer an adequate description of the system. 
Alternatively, one could create a dark soliton in a weakly-interacting 1D BEC, and
gradually increase the effective interaction strength, as done in some experiments.
In this case, one should be able to observe 
a gradual monotonic increase of the oscillation frequency, 
even though the longitudinal confinement remains unaffected.

\section{Conclusions}

In summary, we have discussed  the dynamics of 
both linear and nonlinear excitations within a generalized nonlinear Schr\"{o}dinger equation motivated
by considerations of the behaviour of ultracold atomic 1D Bose gases. The considered model differs from relevant ones 
appearing in the context of optics \cite{book} 
in that the linear behaviour in the density 
dominates at high densities, with quadratic behaviour in the density 
dominating at lower densities.
Within this generalised model, discussed here for the first time in relation to dark solitons in 1D ultracold Bose gases, 
we have studied the dynamics of dark solitons
%
in the ``crossover regime'' in the effective atomic interaction strength, which is approximately marked by a critical density. 
Our main conclusion, stemming from analytical considerations and
confirmed by numerics, is that the soliton oscillation frequency in this regime lies between the known values arising 
in the two limiting cases of weakly and strongly interacting gases, indicating a continuous
change between these two regimes. 
Finally, we note that although motivated by a particular
physical system, our model and analysis are quite general and are not 
restricted to the details of this particular system.

\acknowledgments
It is a pleasure to acknowledge discussions with Luis Santos and Joachim Brand on the form of
the generalized nonlinearity, and with Vladimir Konotop regarding the applicability of the 
analytical approach. The work of D.J.F. was partially supported by the Special Research Account 
of the University of Athens.
PGK gratefully acknowledges support from NSF-DMS-0204585, 
NSF-DMS-0505663 and NSF-CAREER.



\end{document}